\documentclass{nature}
\usepackage{graphicx}
\usepackage{placeins}
\usepackage{nature_journal_style}
\usepackage{deluxetable}
\usepackage{hyperref}

\usepackage{datetime}
\usepackage{amsmath}

\usepackage[utf8]{inputenc}

\usepackage[ngerman, num]{isodate}
\monthyearsepgerman{\,}{\,}

\usepackage{color}
\definecolor{shadecolor}{rgb}{0.9,0.9,0.9}

\title{
  \begin{center}
    \colorbox{shadecolor}{Author's version, 
      Nature BCA, 12thSept.2018}
  \end{center}
Does the galaxy \emph{NGC1052-DF2} falsify Milgromian dynamics? 
 }
\author{Pavel~Kroupa$^{1,2}$, Hosein Haghi$^{1,3}$, Behnam Javanmardi$^{4}$, Akram
  Hasani Zonoozi$^{1,3}$, Oliver M\"uller$^{5}$, Indranil Banik$^6$,
  Xufen Wu$^{7,8}$, Hongsheng Zhao$^{6,9}$, J\"org Dabringhausen$^2$}
\begin{document}

\maketitle

\begin{affiliations}
\item Helmholtz-Institut f\"ur Strahlen- und Kernphysik, University
  of Bonn, Nussallee 14-16, D-53115 Bonn, Germany
\item Charles University in Prague, Faculty of Mathematics and
  Physics, Astronomical Institute, V Hole\v{s}ovi\v{c}k\'ach 2, CZ-180
  00 Praha 8, Czech Republic
\item Institute for Advanced Studies in Basic Sciences, Physics
  department, Gavazang road, Zanjan, 4513766731 Zanjan, Iran
\item School of Astronomy,
Institute for Research in Fundamental Sciences (IPM)
P. O. Box 19395-5531, Tehran, Iran
\item Departement Physik, Universität Basel, Klingelbergstrasse 82,
 CH-4056 Basel, Switzerland
\item Scottish Universities Physics Alliance, University of St
  Andrews, North Haugh, St Andrews, Fife KY16 9SS, UK 
\item CAS Key Laboratory for Research in Galaxies and Cosmology, Department of 
Astronomy, University of Science and Technology of China, Hefei, 230026, P.R. China
\item School of Astronomy and Space Science, University of Science and Technology 
of China, Hefei 230026, China
\item Department of Physics and Astronomy, LaserLaB, Vrije Universiteit, De Boelelaan
  1081, NL-1081 HV Amsterdam, the Netherlands
\end{affiliations}

\clearpage
\textbf{A great challenge in present-day physics is to understand
  whether the observed internal dynamics of galaxies is due to dark
  matter matter or due to a modification of the law of
  gravity. Recently, van Dokkum et al.\cite{vanDokkum+18} reported
  that the ultra-diffuse dwarf galaxy NGC1052-DF2 lacks dark matter,
  and they claimed that this would – paradoxically – be problematic
  for modified gravity theories like Milgromian dynamics
  (MOND\cite{Milgrom83, FM12}). However, NGC1052-DF2 is not isolated,
  so that a valid prediction of its internal dynamics in MOND cannot
  be made without properly accounting for the external gravitational
  fields from neighbouring galaxies. Including this external field
  effect following Haghi et al.\cite{Haghi09} shows that NGC1052-DF2
  is consistent with MOND.  }

In any viable cosmological model, both primordial and tidal dwarf
galaxies, which form in gas-rich tidal debris when galaxies interact,
should exist. Within the standard dark-matter based cosmological
model, primordial dwarfs are dark-matter dominated, whereas tidal
dwarf galaxies contain very little (if any) dark
matter\cite{Ploeckinger+18}. In MOND -- a classical potential theory
of gravity derivable from a Lagrangian with conserved energy, total
momentum and angular momentum \cite{Milgrom83, BM84} -- the two types
of dwarf galaxies cannot be distinguished.  Until now, all known dwarf
galaxies have shown similar dynamical behaviour, possibly implying
falsification of the dark-matter models\cite{Kroupa12a}. The discovery
by van Dokkum et al.\cite{vanDokkum+18} of a galaxy lacking dark
matter would thus constitute an important verification of the standard
dark-matter cosmological model.

Given the density distribution of baryonic matter, $\rho$, the
gravitational potential of Milgromian gravitation, $\phi$, is
determined by the generalised nonlinear Poisson equation\cite{BM84}
$\vec{\nabla} \cdot \left[ \mu \left( | \vec{\nabla}\phi |/a_o \right)
  \vec{\nabla} \phi \right] = 4\,\pi\,G\,\rho$, where the function
$\mu \left( | \vec{\nabla}\phi |/a_o \right)$ describes the transition
from the Newtonian ($| \vec{\nabla}\phi |/a_o \gg 1$) to the
Milgromian ($| \vec{\nabla}\phi |/a_o \ll 1$)
regime\cite{FM12,Milgrom99}, $a_0 = 3600\,$km$^2$s$^{-2}$kpc$^{-1}$ is
Milgrom's constant\cite{Milgrom83,FM12} and $G$ is Newton's universal
constant of gravitation.  An equation of this form has been used in
investigations of classical models of quark dynamics and is therefore
not without precedent in physics\cite{BM84}. One implication of this
equation is that the coupling strength of an object described by
$\rho$ depends on the position of another object, whereas its inertial
mass is given only by its baryonic component. This breaking of the
equivalence of the active gravitating and inertial mass of an object
constitutes a prediction of a new physical phenomenon, which does not
exist in standard gravitation models and may be visualised as the
phantom dark-matter halo (incorporated in $\phi$) being reduced in the
presence of a sufficiently strong constant external
field\cite{Milgrom1986}. This may be related to the quantum
vacuum\cite{Milgrom99}. A dwarf galaxy in the vicinity of a major
galaxy may thus lose its phantom dark-matter halo appearing as a
purely Newtonian system\cite{FM12, McGaugh16a, Thomas+18}.

Following van Dokkum et al.\cite{vanDokkum+18} we assume
\emph{NGC1052-DF2} is located at a distance of $D=20\,$Mpc from the
Earth in the \emph{NGC1052} group, has a globular cluster population
with an effective radius of $r_e = 3.1\,$kpc (the stellar body has
$r_e\approx 2.2\,$kpc), an absolute V-band magnitde $M_V = -15.4$ and
a mass-to-light ratio of the stellar population of
$M_{\rm DF2}/L_V=2$, where $M_{\rm DF2}$ is the baryonic mass of
\emph{NGC1052-DF2} and $L_V$ is its absolute luminosity in the visual
photometric band.
If \emph{NGC1052-DF2} were isolated, then its MOND internal
acceleration would be
$a_i=570\,{\rm km}^2 {\rm s}^{-2} {\rm kpc}^{-1 }= 0.15\,a_0$.  This
weak gravity would imply a MONDian velocity dispersion of
$\sigma_{\rm M} \approx \left( \left(4/81\right) \,G\,M_{\rm
    DF2}\,a_o\right)^{1/4}\approx 20\,$km/s.  The $2\sigma$
($3\sigma$) confidence range for the observed velocity dispersion of
the globular clusters, calculated using the robust biweight
estimator\cite{Beers+90}, is $0.0 - 14.2\,$km/s ($0.0 - 19.7\,$km/s),
falsifying MOND with high significance\cite{vanDokkum+18}.

This inference is only correct if \emph{NGC1052-DF2} is sufficiently
isolated and at the distance of $D\approx 20\,$Mpc.  The bright
elliptical host galaxy \emph{NGC1052} has a baryonic mass\cite{
  Fernandez-Ontiveros+11, Bellstedt+18}
$M_{\rm NGC1052} \approx 10^{11} \, M_\odot$.  The projected
separation between \emph{NGC1052-DF2} and \emph{NGC1052} is
$D_{\rm sep} = 80\,$kpc if both are at the same distance of
$D = 20\,$Mpc. At this separation, the acceleration induced by the
MOND external-field (EF) of the host,
$a_e = \sqrt{a_0\times G \times M_{\rm NGC1052}} / D_{\rm sep}$, is
$a_e \approx 0.14\,a_0$, such that $a_i \approx a_e \ll a_0$. Because
the accelaration by the external-field is as important as internal
galactic acceleration, it needs to be properly taken into account in
calculating the velocity dispersion in MOND, $\sigma_{\rm M}$.  This
is true for any object of mass $m$ whenever its physical size
$r_e > (m/M)^{1/2} \, D_{\rm sep}$ and the external field is caused by
another object of mass $M$ at a distance of $D_{\rm sep}$.

\begin{figure}
\begin{center}
\includegraphics[width=12cm]{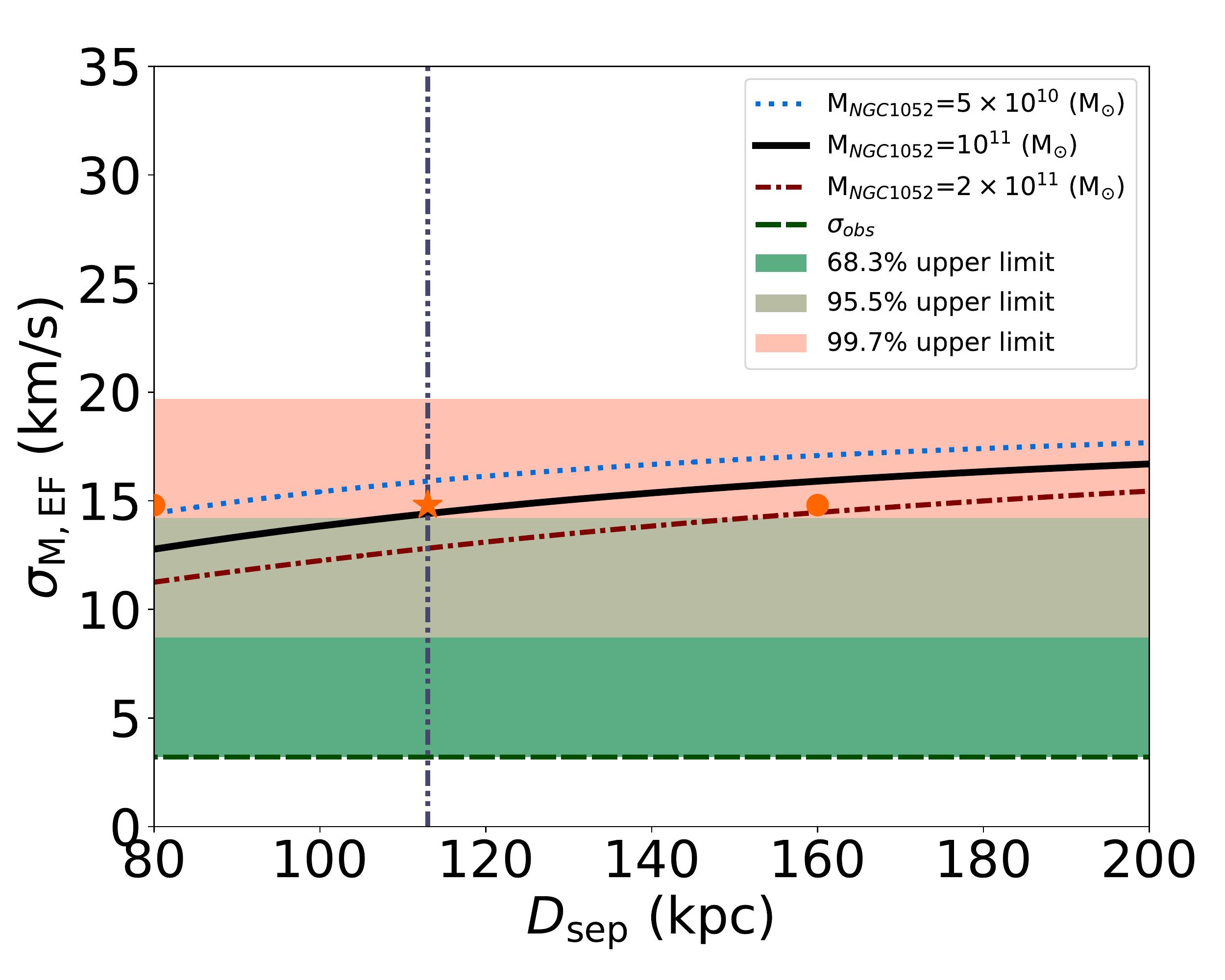}
\end{center}
\vspace{-15mm}
\caption{\bf MOND velocity dispersion and external field} {\small The
  line-of-sight MOND velocity dispersion, $\sigma_{\rm M, EF}$, of
  \emph{NGC1052-DF2} is shown as a function of the separation,
  $D_{\rm sep}$, between \emph{NGC1052-DF2} and \emph{NGC1052}. The
  star (at $D_{\rm sep}=113\,$kpc,
  $M_{\rm NGC1052} \approx 10^{11}\,M_\odot$) and the two orange
  points (at $D_{\rm sep}=80\,$kpc,
  $M_{\rm NGC1052} \approx 5\times 10^{10}\,M_\odot$ and at
  $160\,$kpc, $M_{\rm NGC1052} \approx 2\times10^{11}\,M_\odot$)
  correspond to the $a_e = 0.1 \, a_0$ entry in table~1 of Haghi et
  al. (2009)\cite{Haghi09} without interpolation.  The vertical
  dash-dotted line indicates the more likely distance
  $D_{\rm sep}=\left( 2 \right) ^{1/2} 80\, = 113$kpc.  The $1\sigma$,
  $2\sigma$ and $3\sigma$ upper confidence limits on the observed
  velocity dispersion\cite{vanDokkum+18},
  $\sigma_{\rm obs}=3.2^{+5.5}_{-3.2}\,$km/s, are coloured as
  indicated in the key. The measured value is shown as the long-dashed
  horizontal line.  Our interpolation of the values in table~1 of
  Haghi et al.\cite{Haghi09} (equation~\ref{eq:EF} in Supplementary
  Information) is shown for
  $M_{\rm NGC1052}=5 \times 10^{10}\,M_\odot, 1 \times
  10^{11}\,M_\odot$ and $2\times 10^{11}\,M_\odot$ by the dotted,
  solid and dash-dotted lines, respectively.  }
\label{fig:MOND}
\end{figure}

Using a MOND N-body integrator\cite{NMODY}, Haghi et al.\cite{Haghi09}
quantified the global 1-D line-of-sight velocity dispersion
($\sigma_{\rm M, EF}$) of a non-isolated stellar system lying in the
intermediate external-field regime (see Supplementary Information for
more details).  The dependence of the velocity dispersion of
\emph{NGC1052-DF2}, $\sigma_{\rm M, EF}$, on $D_{\rm sep}$ and
$M_{\rm NGC1052}$ is shown in Fig.~\ref{fig:MOND}.  The velocity
dispersion declines from the isolated, asymptotic value
$\sigma_{\rm M} = 20\,$km/s with decreasing $D_{\rm sep}$ and
increasing mass because the external field of \emph{NGC1052}
suppresses the phantom dark-matter halo, leading to Newtonian
behaviour in the case of $a_e \gg a_i$.  For the nominal host
mass\cite{Bellstedt+18}, $M_{\rm NGC1052}=10^{11}\,M_\odot$ and for
$D_{\rm sep}=113\,$kpc, the MOND velocity dispersion is in agreement
within the $2\sigma$ confidence range of the observational value of
van Dokkum et al.\cite{vanDokkum+18}, as shown by the star in
Fig.~\ref{fig:MOND}.  Thus, the estimate of the internal velocity
dispersion of the globular cluster system of \emph{NGC1052-DF2}
according to MOND is $\sigma_{\rm M, EF} = 14\,$km/s for
$M_{\rm NGC1052}=10^{11}\,M_\odot$ and $D_{\rm sep}=113\,$kpc
(Fig.~\ref{fig:MOND}), in agreement with the observed value within
the~$2\sigma$ confidence level.  \emph{NGC1052-DF2} is thus consistent
with an important MOND prediction.

For the external-field to suppress the phantom dark-matter halo of
\emph{NGC1052-DF2}, this galaxy needs to be close to
\emph{NGC1052}. In ref.\cite{vanDokkum+18} the distance of
\emph{NGC1052-DF2} is estimated using the surface-brightness
fluctuation method.
The calculations performed here show that \emph{NGC1052-DF2} falsifies
MOND if its distance from Earth is $D > 18\,$Mpc and its physical
separation from \emph{NGC 1052} or any other massive galaxy is
$D_{\rm sep} > 300\,$kpc. In this case, the external field becomes
negligible and the galaxy has too small a velocity dispersion for its
stellar mass in MOND. On the other hand, the observed apparent
luminosity of the galaxy is $l_V = L_V / D^2$. In Milgromian dynamics,
$\sigma_{\rm M} \propto M_{\rm DF2}^{1/4} \propto L_V^{1/4} \propto
D^{1/2}$, so \emph{NGC1052-DF2} would be consistent with MOND at the
$2\sigma$ confidence level if $D<13\,$Mpc even if it were isolated.
We note that the velocity dispersion predicted by Newtonian dynamics
is also proportional to $D^{1/2}$ for a galaxy without dark matter.
Future high-precision observations will ascertain the distance of
\emph{NGC1052-DF2} and will constrain its gravitating mass using
stellar velocities within \emph{NGC1052-DF2}, instead of globular
cluster velocities.  As a word of caution, if this galaxy is not in
virial equilibrium then using the globular clusters as tracers for the
mass may be unreliable, independently of the underlying cosmological
model.

\noindent{\bf Online Supplementary Information} contains the external
field effect calculation in MOND.

\clearpage

\noindent{\bf References}
\bibliographystyle{naturemag}
\bibliography{/Users/pavel/PAPERS/BIBL_REFERENCES/kroupa_ref}

\clearpage

\noindent{\bf Acknowledgements}

\noindent 
We acknowledge support from the Alexander von Humboldt Foundation, the
German Academic Exchange Service (DAAD), the Swiss National Science
Foundation, the Natural Science Foundation of China (grant numbers
11503025 and 11421303), the Anhui Natural Science Foundation (grant
number 1708085MA20), the 'Hundred Talents Project of Anhui Province',
the 'Fundamental Research Funds for the Central Universities', the
'DAAD- Ostpartnerschaftsprogramm f\"ur 2018' and the Iran Science Elites
Federation (ISEF).

\noindent{\bf Author contributions:}

\noindent P.K. conceived the work and wrote the mansucript; H.H. and
A.H.Z. performed the external field calculations; O.M.  and
B.J. contributed observational expertise on galaxies; I.B., X.W.  and
H.Z. contributed MOND expertise; J.D.  contributed dwarf galaxy
expertise. All co-authors contributed to the manuscript with comments
and passages.

\noindent{\bf Statement on competing interests:}

\noindent Declared none.

\clearpage

\clearpage

\noindent{\bf Supplementary Information}

\vspace{5mm}

\noindent {\bf The external field effect in MOND:}
In a self-bound gravitational system with baryonic mass $M$ and
half-mass radius $r_h$, its components experience a Newtonian internal
acceleration, $a_N \equiv G\,M/\left( 2\,r_h^2\right)$, which is in
Newtonian gravitation linearly additive to any external acceleration
$a_e$. If this external acceleration is constant across the system
then it does not influence the internal eigengravitation.  However, in
MOND, the internal gravitation is suppressed by the external field if
it is sufficiently strong.  This is known since the original work by
Milgrom (1983, ApJ 270, 365)\cite{Milgrom83} and follows from the
Lagrangian formulation of MOND which implies the generalised
non-linear Poisson equation (Bekenstein \& Milgrom 1984, ApJ 286,
7)\cite{BM84}.  The details of computing the acceleration on the
systems's constituents are simple only in the limits where the
internal acceleration is much larger than Milgrom's constant $a_0$ (in
which case the system is Newtonian), or when the external field is
negligible and the internal acceleration is much smaller than
$a_0$. In the latter case the system is in the deep MOND limit such
that its MOND internal acceleration is
$a_i = \sqrt{(G\, M\, a_0)}/r_e$ for a system with effective radius
$r_e$, and effectively behaves as if it is embedded in a dark-matter
halo which can be described mathematically as a Newtonian isothermal
logarithmic potential. This Milgromian potential is referred to as a
phantom dark-matter halo as it is not made of particles, and this
limiting case was used by van Dokkum et al.\cite{vanDokkum+18} to
calculate the value of the velocity dispersion predicted by MOND.  In
the case when the external field is significant compared to the
internal acceleration and when both are smaller than or comparable to
Milgrom's constant $a_0$ then the calculations become less
straightforward, and again, only in the limiting cases can
calculations be performed.  For example, when $a_e > a_0$ and
$a_i \ll a_ e$ forces from different masses can be superposed (Banik
\& Zhao 2015, arXiv150908457B), though this is generally not correct
as MOND is non-linear.  Effectively an external field truncates (or
reduces or ``melts'') the phantom dark-matter halo of the system.  The
external field effect thereby allows observational tests for MOND,
because it has no counterpart in Newtonian dynamics.  Simplified
descriptions of the external-field effect are available in Wu \&
Kroupa (2015, MNRAS 446, 330) and an in-depth discussion is found in
the major review by Famaey \& McGaugh (2012)\cite{FM12}.

Observational evidence for the MOND external-field effect has been
found in the form of falling rotation curves of disk galaxies (Wu \&
Kroupa 2015, MNRAS, 446, 330; Hees et al. 2016, MNRAS, 455, 449; Haghi
et al. 2016, MNRAS 458, 4172), and in the asymmetric tidal tail of the
globular star cluster Pal~5 (Thomas et al. 2018\cite{Thomas+18}).  The
successful prediction of the very low velocity dispersion of the
ultra-diffuse dwarf satellite galaxy Crater~II\cite{McGaugh16a} and the
velocity dispersions of Andromeda satellite galaxies (McGaugh \&
Milgrom 2013, ApJ 775, 139) are highly relevant for the problem at
hand.

Haghi et al.\cite{Haghi09} quantified the global 1-D line-of-sight
velocity dispersion ($\sigma_{\rm M, EF}$ ) of a non-isolated stellar
system lying in the intermediate external-field regime, using the code
{N-MODY}, which has a N-body MOND mean-field potential solver
developed by Ciotti, Londrillo \& Nipoti (2006, ApJ 640,
741)\cite{NMODY}.  Haghi et al.\cite{Haghi09} fit their results for
$\sigma_{\rm M, EF}$ in the intermediate regime
($a_i \approx a_e < a_0$) for different values of the external
field. They quantified the different asymptotic behavior in fig. 5 of
Haghi et al.\cite{Haghi09} (i.e. in the Newtonian, the deep-MOND and
and the external-field-dominated regime). The formulation was
presented as their eq.~16 and~17 with coefficients provided for
different values of $a_e$ in their table~1.

Thus, for example, the $M=10^{11}\,M_\odot$,
$D_{\rm sep} =\sqrt{2} \times 80 = 113\,$kpc case corresponds to
$a_e = 0.1\,a_0$.  Adopting the three coefficients tabulated in the
row $a_e=0.1\,a_0$  in table~1 of Haghi et al. (2009)\cite{Haghi09},
we use their eqs.~13, 15 and 17, to obtain $\sigma = 14.8\,$km/s.  This
is indicated by the star symbol in Fig.~\ref{fig:MOND}.  Likewise for
other values of $D_{\rm sep}$ and $M$ combinations, we can work out
$a_e$, and the corresponding $\sigma_{M, EF}$, and these are plotted as curves
in Fig.~\ref{fig:MOND}.  

{\it The interpolation to other-than-the tabulated values of $a_e$ in
table~1 in Haghi et al. (2009)}\cite{Haghi09} is not unique and for the
purpose of documentation we provide in the following one possible
analytical formulation (eq.~\ref{eq:EF}) for the data in their table~1
allowing $\sigma_{\rm M, EF}$ to be calculated as a function of the
internal field in the system with mass $M$ exposed to an external
field, $a_e$.  The one-dimensional (line of sight) velocity
dispersion, $\sigma_{\rm M, EF}$ (in km/s),
\begin{equation}
{\rm log}_{10} \sigma_{\rm M, EF} = {\rm log}_{10} \sigma_{\rm M} + F(a_e),
\label{eq:EF}
\end{equation}
where the velocity dispersion of an isolated system
\begin{equation}
\sigma_{\rm M} =
\left( {4\over 81}\, G\, M\, a_0 \right)^{1\over4} \times
\left(1 +
0.56\,{\rm exp} \left( 3.02 \, x \right)
\right)^{0.184},
\label{eq:isol}
\end{equation}
where $x \equiv {\rm log}_{10}\left( a_N / a_0 \right)$ and
\begin{equation}
F(a_e) = - { A(a_e) \over 4} \,
\left( {\rm ln}  \left[ {\rm exp}
\left( -{x\over A(a_e)} \right) + B(a_e) \right] + C(a_e) \right),
\end{equation}
with $G$ being Newton's gravitational constant. Note that
Eq.~\ref{eq:isol} supersedes the equation used by van Dokkum et
al.\cite{vanDokkum+18} by a correction factor which ensures the
correct behaviour as the internal acceleration approaches the
Newtonian value.

The functions $A, B, C$ are fits to the data in table~1 in Haghi et al.\cite{Haghi09}:
\begin{align}
A(a_e) &= {5.3 \over
\left(
10.56 + \left( y +2 \right)^{3.22}
\right)  }, \\
B(a_e) &=
0.44 \, {\rm exp} \left(-4.26\, y \right),\\
C(a_e) &= -12.83\,
{\rm ln}
\left(
{\rm exp} \left(  {-\left( y + 2 \right)  \over 1.28} \right)
 + 0.85
\right),
\end{align}
where $y \equiv {\rm log}_{10} \left( a_e/a_0 \right)$. The important
issue is to obtain an accurate fit to the combination of $A$, $B$ and
$C$ given in eq.~15 of Haghi et al. (2009)\cite{Haghi09} $-$ fitting
the coefficients individually is less important.

These formulae reproduce well the previous analytical velocity
dispersion estimators in systems 
in the isolated deep-MOND regime [that is,
$\sigma_{\rm M, EF} \approx \sigma_{M} = \left( 4\,G\,M \, a_0 / 81
\right)^{1/4}$] (Milgrom, M. 1994, ApJ 429, 540; McGaugh, S. \&
Milgrom, M., 2013, ApJ 775, 139) and for the deep-MOND
external-field-dominated regime
($\sigma_{\rm M, EF} \propto 1/\sqrt{(a_e)}$, eq.~2 of McGaugh
2016\cite{McGaugh16a}).
We note that the three orange points in Fig.~\ref{fig:MOND} can be obtained
directly from the $a_e = 0.1 \, a_0$ line in table~1 of Haghi et
al. (2009)\cite{Haghi09} without interpolation.

\end{document}